\documentclass[twoside,twocolumn,9pt]{article}
\usepackage{extsizes}
\usepackage[super,sort&compress,comma]{natbib} 
\usepackage[version=3]{mhchem}
\usepackage[left=1.5cm, right=1.5cm, top=1.785cm, bottom=2.0cm]{geometry}
\usepackage{balance}
\usepackage{mathptmx}
\usepackage{sectsty}
\usepackage{graphicx} 
\usepackage{lastpage}
\usepackage[format=plain,justification=justified,singlelinecheck=false,font={stretch=1.125,small,sf},labelfont=bf,labelsep=space]{caption}
\usepackage{float}
\usepackage{fancyhdr}
\usepackage{fnpos}
\usepackage[english]{babel}
\addto{\captionsenglish}{%
  
}
\usepackage{array}
\usepackage{droidsans}
\usepackage{charter}
\usepackage[T1]{fontenc}
\usepackage[usenames,dvipsnames]{xcolor}
\usepackage{setspace}
\usepackage[compact]{titlesec}
\usepackage{hyperref}

\usepackage{epstopdf}

\definecolor{cream}{RGB}{222,217,201}

\newcommand{\etal}{\textit{et al}.\ }

\newcommand{\eg}{\textit{e}.\textit{g}.\ }

\begin{document}

\pagestyle{fancy}
\thispagestyle{plain}
\fancypagestyle{plain}{
\renewcommand{\headrulewidth}{0pt}
}

\makeFNbottom
\makeatletter
\renewcommand\LARGE{\@setfontsize\LARGE{15pt}{17}}
\renewcommand\Large{\@setfontsize\Large{12pt}{14}}
\renewcommand\large{\@setfontsize\large{10pt}{12}}
\renewcommand\footnotesize{\@setfontsize\footnotesize{7pt}{10}}
\makeatother

\renewcommand{\thefootnote}{\fnsymbol{footnote}}
\renewcommand\footnoterule{\vspace*{1pt}%
\color{cream}\hrule width 3.5in height 0.4pt \color{black}\vspace*{5pt}} 
\setcounter{secnumdepth}{5}

\makeatletter 
\renewcommand\@biblabel[1]{#1}            
\renewcommand\@makefntext[1]%
{\noindent\makebox[0pt][r]{\@thefnmark\,}#1}
\makeatother 
\renewcommand{\figurename}{\small{Fig.}~}
\sectionfont{\sffamily\Large}
\subsectionfont{\normalsize}
\subsubsectionfont{\bf}
\setstretch{1.125} 
\setlength{\skip\footins}{0.8cm}
\setlength{\footnotesep}{0.25cm}
\setlength{\jot}{10pt}
\titlespacing*{\section}{0pt}{4pt}{4pt}
\titlespacing*{\subsection}{0pt}{15pt}{1pt}

\fancyfoot{}
\fancyfoot[LO,RE]{\vspace{-7.1pt}\includegraphics[height=9pt]{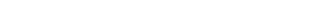}}
\fancyfoot[CO]{\vspace{-7.1pt}\hspace{13.2cm}\includegraphics{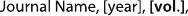}}
\fancyfoot[CE]{\vspace{-7.2pt}\hspace{-14.2cm}\includegraphics{head_foot/RF}}
\fancyfoot[RO]{\footnotesize{\sffamily{1--\pageref{LastPage} ~\textbar  \hspace{2pt}\thepage}}}
\fancyfoot[LE]{\footnotesize{\sffamily{\thepage~\textbar\hspace{3.45cm} 1--\pageref{LastPage}}}}
\fancyhead{}
\renewcommand{\headrulewidth}{0pt} 
\renewcommand{\footrulewidth}{0pt}
\setlength{\arrayrulewidth}{1pt}
\setlength{\columnsep}{6.5mm}
\setlength\bibsep{1pt}

\makeatletter 
\newlength{\figrulesep} 
\setlength{\figrulesep}{0.5\textfloatsep} 

\newcommand{\topfigrule}{\vspace*{-1pt}%
\noindent{\color{cream}\rule[-\figrulesep]{\columnwidth}{1.5pt}} }

\newcommand{\botfigrule}{\vspace*{-2pt}%
\noindent{\color{cream}\rule[\figrulesep]{\columnwidth}{1.5pt}} }

\newcommand{\dblfigrule}{\vspace*{-1pt}%
\noindent{\color{cream}\rule[-\figrulesep]{\textwidth}{1.5pt}} }

\makeatother

\twocolumn[
  \begin{@twocolumnfalse}
{\includegraphics[height=30pt]{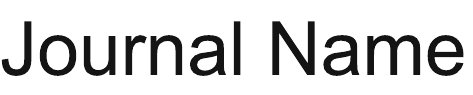}\hfill\raisebox{0pt}[0pt][0pt]{\includegraphics[height=55pt]{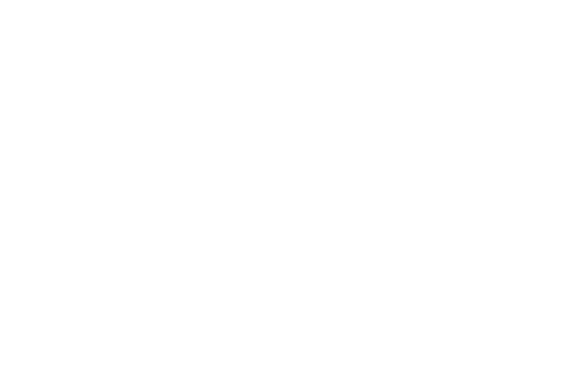}}\\[1ex]
\includegraphics[width=18.5cm]{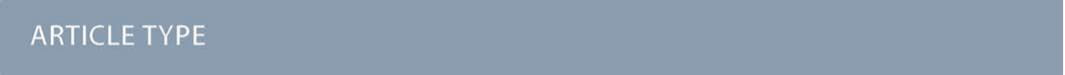}}\par
\vspace{1em}
\sffamily
\begin{tabular}{m{4.5cm} p{13.5cm} }

\includegraphics{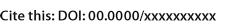} & \noindent\LARGE{\textbf{Lateral Graphene-Metallene Interfaces at the Nanoscale$^\dag$}} \\
\vspace{0.3cm} & \vspace{0.3cm} \\

 & \noindent\large{Mohammad Bagheri,\textit{$^{a}$} and Pekka Koskinen,$^{\ast}$\textit{$^{a}$}} \\

\includegraphics{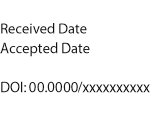} & \noindent\normalsize{Metallenes are atomically thin, nonlayered two-dimensional materials. While they have appealing properties, their isotropic metallic bonding makes their stabilization difficult and presents considerable challenges to their synthesis and practical applications. However, their stabilization can still be achieved by suspending them in the pores of two-dimensional template materials, making the properties of lateral interfaces of metallenes scientifically relevant. Here, we combined density-functional theory and universal machine-learning interatomic potentials to study lateral interfaces between graphene and 45 metallenes with various profiles. We optimized the interfaces and analyzed their energies, electronic structures, and stabilities at room temperature, defect formations, and structural deformations. While broad trends were identified using machine-learning analysis of all interfaces, density-functional theory was the main tool for studying the microscopic properties of selected elements. We found that the interfaces are the most stable energetically and with respect to lattice mismatch, defect formation, and lateral strain when their profiles were geometrically smooth. The most stable interfaces are found for transition metals. In addition, we demonstrate how universal machine-learning interatomic potentials now offer the accuracy required for the modeling of graphene-metallene interfaces. By systematically expanding the understanding of metallenes' interface properties, we hope these results guide and accelerate their synthesis to enable future applications and benefit from metallenes' appealing properties.} \\

\end{tabular}

 \end{@twocolumnfalse} \vspace{0.6cm}

  ]

\renewcommand*\rmdefault{bch}\normalfont\upshape
\rmfamily
\section*{}
\vspace{-1cm}


\footnotetext{\textit{$^{a}$~Nanoscience Center, Department of Physics, University of Jyv\"askyl\"a, 40014 Jyv\"askyl\"a, Finland; E-mail: pekka.j.koskinen@jyu.fi}}

\footnotetext{\dag~Supplementary Information available: Additional details on designed interfaces and simulations; details on comparisons; validation of the machine-learning potential; and supporting snapshots of selected graphene-metallene interfaces. See DOI: 00.0000/00000000.}



\section{Introduction}
The observation of suspended monolayer iron (carbide) membranes in graphene pores during transmission electron microscopy in 2014 marked a significant milestone in the exploration of two-dimensional (2D) materials \cite{Zhao2014}. Gradually, this experimental observation, which relies on stabilization by lateral interfaces, led to the emergence of a new class of materials: metallenes.
Metallenes are a family of 2D non-van der Waals materials made of atomically thin elemental metals \cite{yang2015glitter, nevalaita2018atlas,prabhu2021metallenes, cao2022ultrathin, Xie2023, shahzad2024recent}.
Their metallic bonding and delocalized electronic structure make them unique among 2D materials and highly attractive for applications in catalysis, sensing, biomedicine, electronics, energy storage, and energy conversion \cite{Lu2023_bio, shahzad2024recent, Xie2023, yaoda2020, Kashiwaya_2025}. However, their isotropic metallic bonding makes their synthesis challenging and hinders their widespread application. Unlike in covalent van der Waals materials, so far the dimensions of successfully synthesized metallenes have been generally limited to the nanometer scale \cite{goldene2024} or to monolayers confined on or inside the pores of 2D templates \cite{Janne2019, Zr_patch, Zhao_Mo, Zhao_2D_gold, Zhu_2D_gold, Ta_Cr, Mendes_Zr, Zhao_2025}. 

Metallene patches in graphene pores have been studied computationally and experimentally \cite{Nevalaita2018, Nevalaita2019, Zr_patch, Zhao_Mo, Zhao_2D_gold, Ta_Cr, Mendes_Zr}. Metal–2D material heterostructures involving graphene, boron nitrides, and transition metal dichalcogenides have been explored \cite{Gong_2010, Xu_2010, Zhang_2015, Wei_2016, Ovando_2019, Avalos_2019, SINGH2023, Pirker_2024, Falorsi_2025}. Both computational and experimental methods have advanced interface design in these systems \cite{Gerber2023, TribChem, Doan_2025}. However, while these interfaces are crucial for stabilizing metallenes, their microscopic structures remain poorly understood. At present, systematic computational studies of interfaces would offer valuable insights for improving the synthesis and stability of metallenes.

Therefore, in this article, we employed density-functional theory (DFT) and universal machine-learning (ML) interatomic potential calculations to study lateral interfaces between graphene and $45$ different metallenes. For a systematic study, we investigate four interface profiles formed by merging zigzag and armchair edges of graphene with straight and staggered edges of a hexagonal metallene lattice and varying lattice mismatch (Fig.~\ref{fig:interfaces}). We optimized the interfaces, calculated their energies, analyzed electronic structures, and tested their stabilities at room temperature, defect formations, and structural deformations. While broad trends were identified by analyzing all constructed interfaces using reliable ML calculations, we used DFT to conduct a deeper investigation of microscopic properties of a selected set of elements (Mg, Cu, Ga, Ag, and Au) from different groups of the periodic table.
At minimal lattice mismatch, our results show interface energies between $0.21\text{--}0.93$ eV/\AA. Interfaces are the most stable when their profiles are smooth, such as with zigzag graphene/straight metallene interfaces. This trend persists even in the presence of defects, under molecular dynamics simulations, and upon stretching of the interface. In contrast, ragged interfaces with poorly matched edges tended to reconstruct.

\section{Materials and Methods}
\subsection{Constructing lateral interfaces}
\begin{figure}[t]
    \centering
    \includegraphics[width=0.8\linewidth]{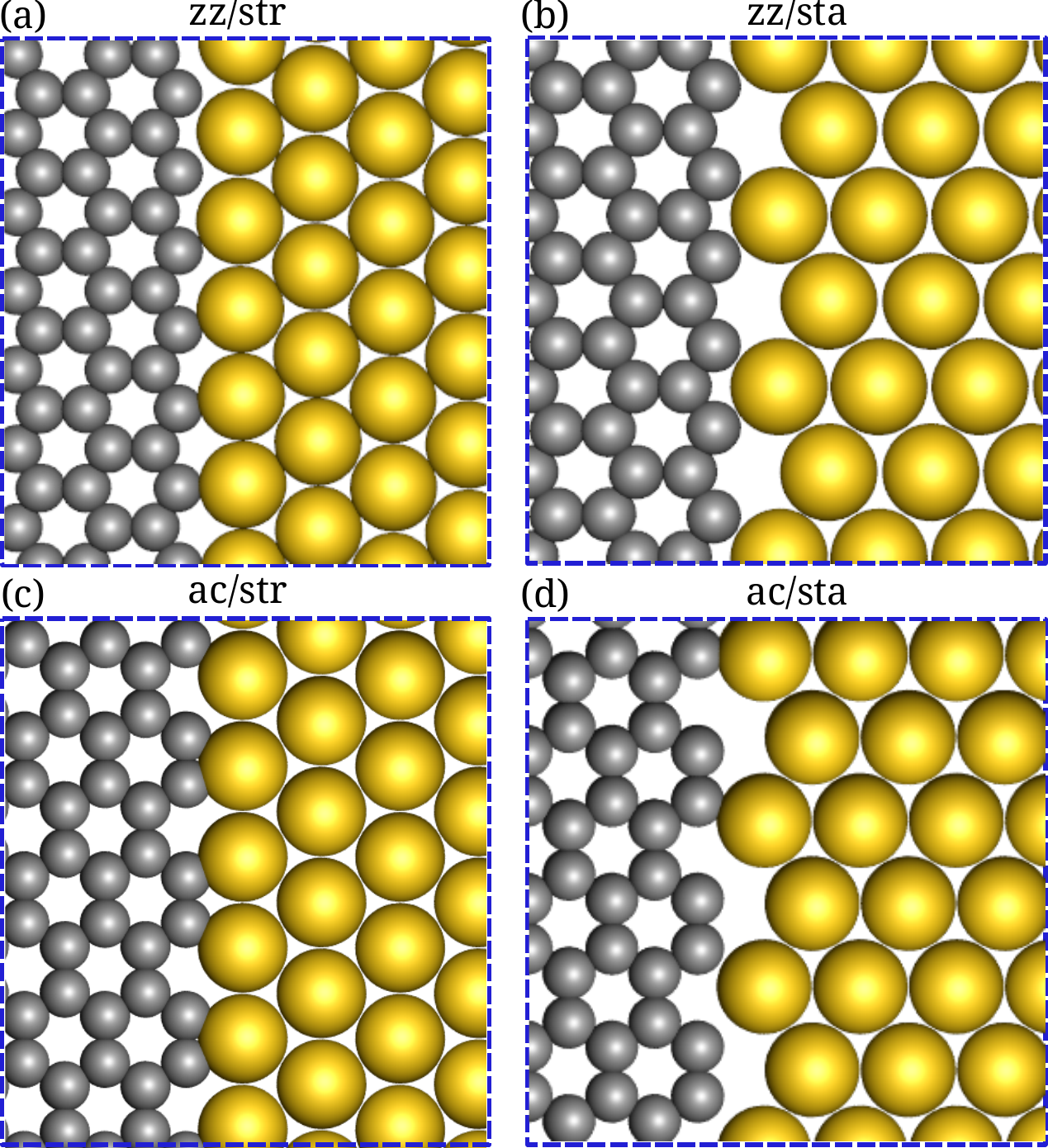}
    \caption{Schematic atomic structures of lateral graphene/metallene interfaces formed by merging zigzag (zz) and armchair (ac) edges of graphene with straight (str) and staggered (sta) edges of hexagonal metallene.
    (a) Graphene zigzag/metallene straight (zz/str), (b) graphene zigzag/metallene staggered (zz/sta), (c) graphene armchair/metallene straight (ac/str), and (d) graphene armchair/metallene staggered (ac/sta) interfaces. Carbon atoms are shown in gray and metal atoms in yellow.
    }
    \label{fig:interfaces}
\end{figure}

We designed the graphene-metallene lateral interfaces by modeling the 2D templates using $7.10$-\AA-wide zigzag (zz) graphene nanoribbons and $8.61$-\AA-wide armchair (ac) graphene nanoribbons. Metallenes were modeled as ribbons of a hexagonal lattice with two edge profiles, straight (str) and staggered (sta). This design results in four types of interface profiles for each metallene (Fig.~\ref{fig:interfaces}). With a slash notation separating the graphene edge/metallene edge, the four edges are zz/str, zz/sta, ac/str, and ac/sta. The practical simulations were performed using a $x$- and $y$-periodic simulation cell with a metallene ribbon of width $w$ and length $l$ confined between the two edges of the graphene nanoribbon (Fig.~S1 in Supporting Information (SI)). Thus, the supercell then consisted of two similar interfaces of length $l$. 

\begin{figure}[!]
    \centering
    \includegraphics[width=\linewidth]{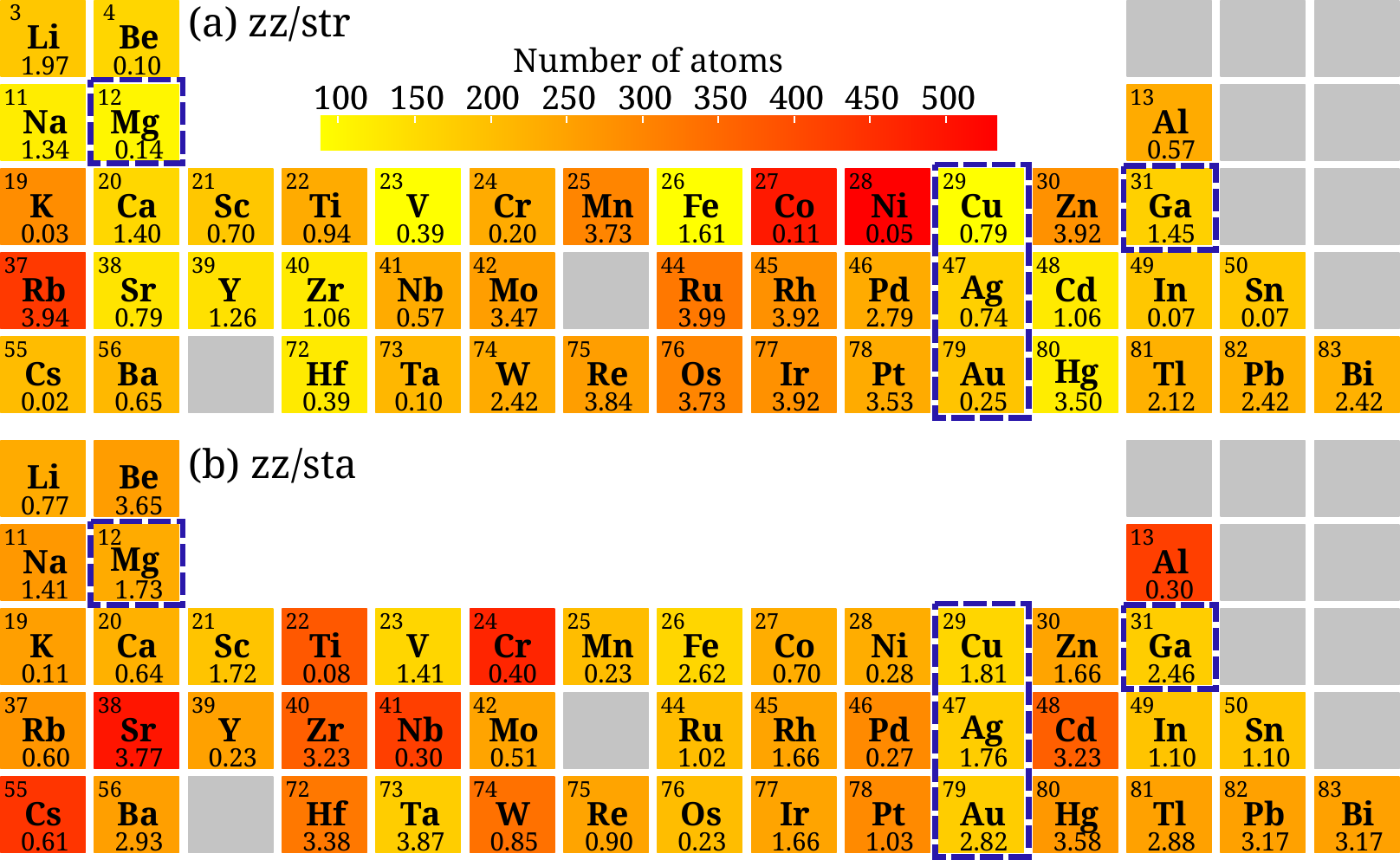}
    \caption{Heat maps of number of atoms for modeled zigzag graphene-metallene interfaces based on selection criteria. a) zigzag/straight and b) zigzag/staggered interfaces.
    The lattice mismatch (strain in graphene) is shown under each element name. The metals selected for DFT calculations are highlighted by dashed boxes.}
    \label{fig:strain-mismatch}
\end{figure}

The vertical lattice mismatch between graphene and metallene was treated by straining the graphene to match the periodic length $l$ of the cell; the metallene ribbon was unstrained with respect to the bond lengths in Ref.~\citenum{nevalaita2018atlas} which are predicted lattice constants of the pristine 2D hexagonal metallenes. For each interface type, one could then choose several lengths $l$ that correspond to various strains in graphene (depending on the number of primitive vertical cells metallene). The aim to minimize the strains usually implies simulation cell sizes that render DFT calculations impractical due to too many atoms. There are $1080$ different graphene/metallene interfaces at lattice mismatch between $0\text{--}5$\% (Figs.~S2 and~S3). Most of the systems with a small strain ($\approx 0$\%) have hundreds to thousands of atoms in the simulation cell. Consequently, our selection criterion was choosing the smallest strain below $3$\%\ while constraining the number of atoms below $250$. The smallest simulation cells for our selection criteria are found mostly for interfaces with graphene zigzag edges (Fig.~\ref{fig:strain-mismatch}). 

Therefore, for a preliminary investigation of optimized lateral interfaces with practical DFT calculations, we chose a handful of elements and focused on graphene(zz)-metallene interfaces (zz/str and zz/sta). Based on Fig.~\ref{fig:strain-mismatch}, we selected five metallenes that had a reasonable number of atoms for both straight and staggered edges and that covered different parts of the periodic table (Table~\ref{table:inf}). Based on previous studies \cite{nevalaita2018atlas, Nevalaita2019, gentle}, apart from Ga, the selected elements rank well regarding the energetic and dynamical stability of gas-phase metallene clusters.
We note that in our design for DFT calculations, the width of metallenes was narrower (4 metal atoms in $x$-direction), which led to a smaller number of atoms in the interfaces. But for most of the elements, the selected width was not enough, and systems deformed during optimization of the structures. So we selected a larger width (6 metal atoms in $x$-direction) for systematic calculations, and we illustrate the number of atoms for the larger width in Figs.~S2 and~S3. We also listed the length of the interfaces ($l$) with minimum lattice mismatch used for ML calculations in Table~S1.

\begin{table}[h]
\small
\caption{\ List of interfaces selected for DFT calculations. The length of the interface is $l$, and the width of the metallene is $w$. The lattice mismatch equals the strain in graphene.}
\label{table:inf}
    \begin{tabular*}{0.48\textwidth}{@{\extracolsep{\fill}}lccc}
    \hline 
        \textbf{Interface} & \textbf{Mismatch (\%)} & \textbf{$w$ (\AA)} & \textbf{$l$ (\AA)} \\ \hline 
        Mg(str)   & 0.1  & 16.0 & 12.3  \\ 
        Mg(sta)   &  1.7 & 26.6 & 26.6 \\  
        Cu(str)   &  0.8 & 12.7 & 9.8  \\ 
        Cu(sta)   & 1.8  & 16.9 & 16.9 \\ 
        Ga(str)   & 1.5  & 14.4 & 19.4 \\ 
        Ga(sta)   & 2.5  & 19.2 & 19.2 \\ 
        Ag(str)   & 0.7  & 14.5 & 19.5 \\ 
        Ag(sta)   & 1.8  & 14.5 & 14.5 \\ 
        Au(str)   & 0.3  & 14.3 & 22.1 \\ 
        Au(sta)   & 2.8  & 9.6 & 9.6 \\ 
        \hline
    \end{tabular*}
\end{table}

\subsection{Computational methods}
The density-functional theory (DFT) calculations were performed in the linear combination of atomic orbitals (LCAO) \cite{lcao} mode of the GPAW code \cite{gpaw1, gpaw3, ase-paper} using the Perdew-Burke-Ernzerhof (PBE) exchange and correlation functional \cite{pbe, PhysRevLett.78.1396, libxc}. The calculations were spin-polarized and converged with respect to vacuum regions, lateral dimensions, and ${\bf k}$-point grids. All calculations had $15$-\AA\ vacuum regions in nonperiodic directions, and Monkhorst-Pack ${\bf k}$-point sampling was $12 \times 12 \times 1$ for the 2D structures and $2 \times 8 \times 1$ for the interfaces \cite{PhysRevB.13.5188, PhysRevB.16.1748}. Under these settings, all structures were relaxed using the BFGS algorithm to forces below $1$~meV/\AA. The atomic energies were obtained from $\Gamma$-point calculations of single atoms in $20$-\AA\ cubic cells. Molecular dynamics (MD) simulations and tensile strengths were calculated in the plane-wave (PW) mode with $800$~eV cutoff energy \cite{pw}. The $1$-ps MD runs used a time step of $2.5$ fs and a room-temperature Langevin thermostat with a friction coefficient of $0.02$~fs$^{-1}$. The PW mode was chosen to directly compare the DFT and the (PW-DFT-trained) machine-learning model, discussed below.

The machine-learning (ML) calculations were performed using MatterSim with a pre-trained model v1.0.0-5M \cite{mattersim}. The MatterSim-v1 model is based on the M3GNet \cite{m3gnet} architecture that was trained on relaxations found in the Materials Project and that has the potential to work across the entire periodic table of the elements \cite{mp}. ML calculations used the same relaxation algorithm, force criteria, and MD parameters as DFT, except for the MD runs that were ten times longer.

The electronic charge transfers were calculated using the algorithm by Henkelman \etal \cite{Tang_2009}, which uses the Bader partitioning scheme to calculate the electronic charging of individual atoms \cite{Bader}.

\section{Results}

\subsection{Interface energies}
\begin{figure}[h]
    \centering
    \includegraphics[width=\linewidth]{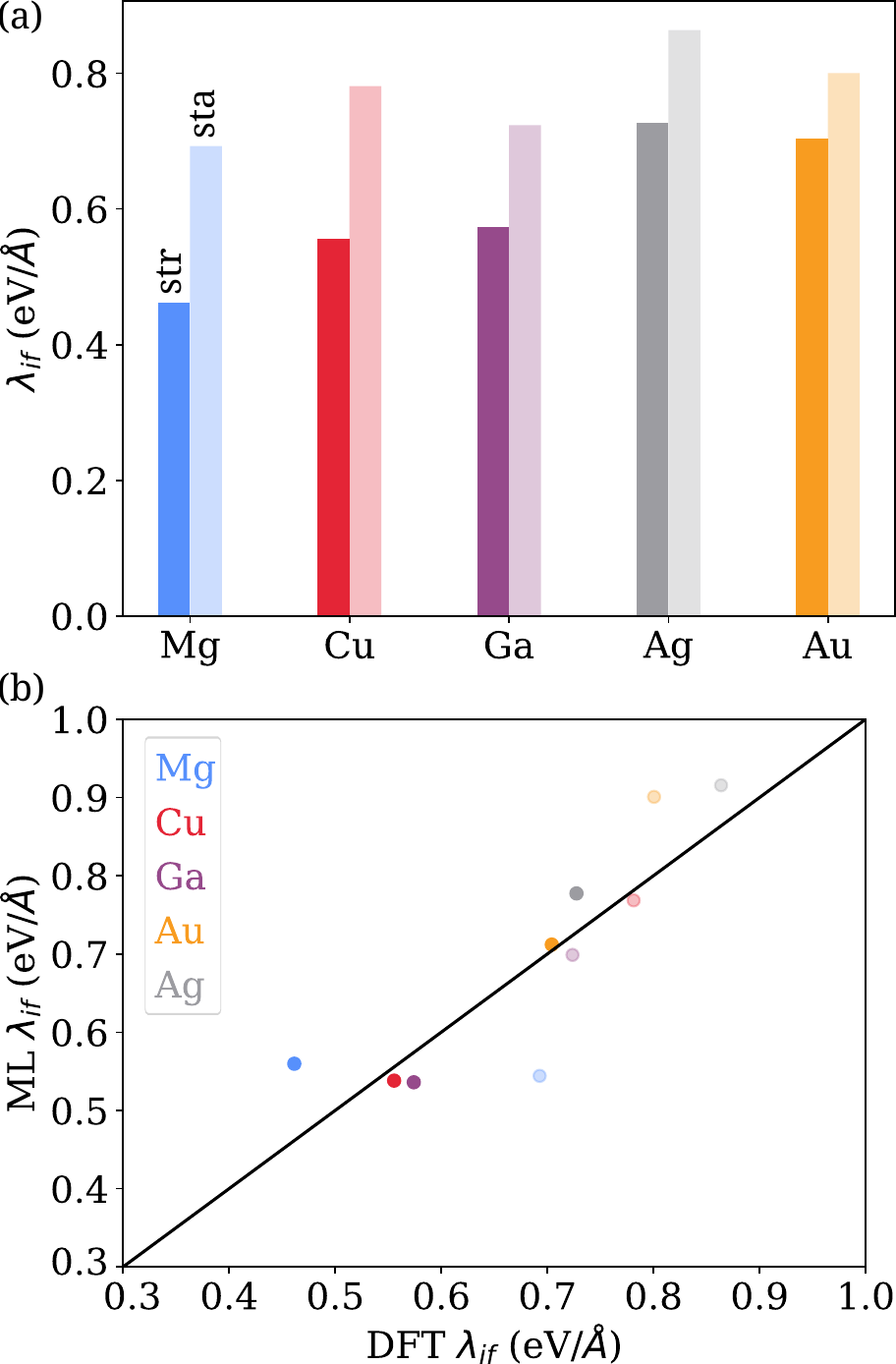}
    \caption{Comparing DFT and ML potentials for interface energies. (a) Calculated interface energies ($\lambda_{if}$) using DFT. (b) Comparison of $\lambda_{if}$ between DFT and ML for the interfaces in Table~\ref{table:inf}. Staggered edges are indicated by a brighter color and straight edges by a darker one.}
    \label{fig:DFT-ML}
\end{figure}

The graphene/metallene interface energy represents the energy required to construct the interface, compared to pristine graphene and metallene without any interfaces. It depends on differences in atomic arrangements, strains, bonding characteristics, charge transfer, and electronic structures \cite{Gao2017, Tea2017, Butler2019, Rizzo2020, Liu2020, Wu2021, Liu2023, Wang2023}. 
We define it as
\begin{equation}
    \lambda_{if} = (E_\text{het} - E_\text{gr} - E_\text{met})/L_\text{if},
    \label{eq:E_inf}
\end{equation}
where $E_\text{het}$ is the energy of heterostructure, $E_\text{gr}$ is the energy of graphene, $E_\text{met}$ is the energy of metallene, and $L_\text{if}=2l$ is the length of interface (Fig.~S1). DFT results show that the interface energies are always larger for the staggered metallene edges (Fig.~\ref{fig:DFT-ML}a). This trend is due to the more ragged interface profile, the concomitant larger nanoscale strains, and the accompanying weaker bonds across the interface \cite{Edge_2025}.

We then compare the DFT results of the same heterostructures to interface energies calculated by ML (Fig.~\ref{fig:DFT-ML}b). There is a good agreement between DFT and ML interface energies, with mean absolute error (MAE) of $0.04$ eV/\AA\ for zz/str and $0.07$ eV/\AA\ for zz/sta interfaces. ML is also capable of modeling the pristine hexagonal metallenes. 
The metallene cohesive energies from ML and DFT (both PW or LCAO) compare reasonably well, although ML overestimates them for some elements (Fig.~S4). ML-predicted energies have MAE of $0.87$~eV for PW and $0.98$~eV for LCAO. Even within DFT, PW and LCAO have relative MAE of $0.32$~eV, with LCAO having larger energy for most of the metals. Despite the small overestimation of ML energies for metallenes, ML captures interface energy trends with fair accuracy. 

Therefore, we can conclude that ML provides a reliable tool to predict structures and energies of graphene/metallene interfaces at DFT accuracy, while enabling the modeling of much larger unit cells and smaller lattice mismatches compared to DFT. We took advantage of this situation and calculated the properties of all the designed zigzag/metallene interfaces. As expected, the resulting interface energies are the smallest at minimum lattice mismatch and increase upon increasing strain (Fig.~S5). Note how the metal and the interface type dominate the interface energies---the strain effect is a mere perturbation. Some exceptions, \eg Co, Mn, Ni, may be caused by the random formation of enhanced bonding across the interface at opportune values of lattice mismatch. Vanadium was excluded from this inspection because it deformed dramatically during structural optimization.

\begin{figure}[t]
    \centering
    \includegraphics[width=\linewidth]{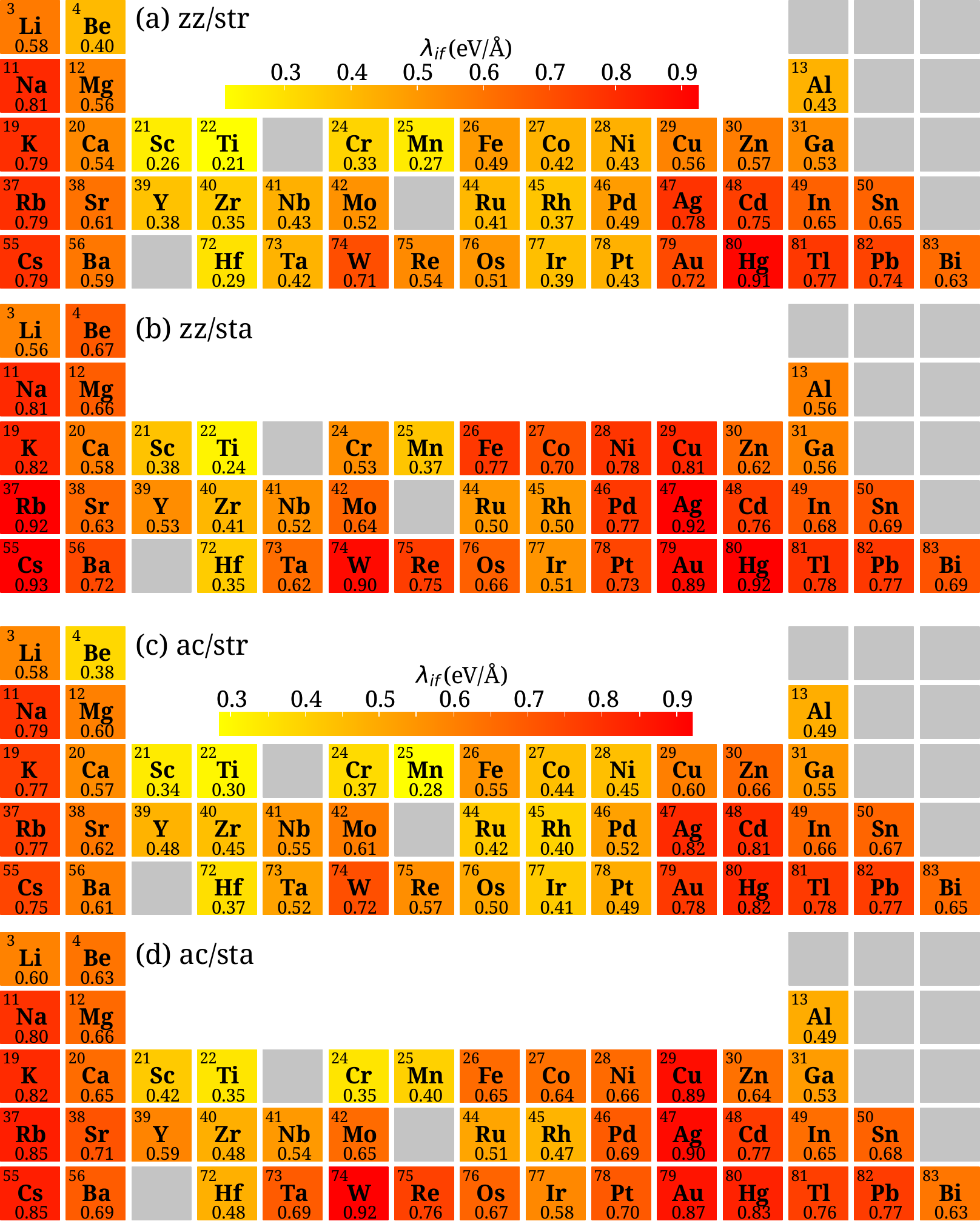}
    \caption{Interface energies. Heat maps of interface energies calculated using ML for a) zigzag/straight, b) zigzag/staggered, c) armchair/straight, and d) armchair/staggered interfaces, corresponding to minimum lattice mismatch. The interface energies are shown under each element.}
    \label{fig:energy_inf}
\end{figure}

Let us then inspect the results of the graphene/metallene interfaces at minimum lattice mismatch. All interface energies are positive, suggesting that graphene and metallene are stable against the spontaneous formation of interfaces (Fig.~\ref{fig:energy_inf}). Early and middle transition metals show lower interface energies than alkali, alkaline earth, and post-transition metals. Higher interface energy indicates that, in relative terms, metal-metal bonds are stronger than carbon-metal bonds. Among zz/str, the interface energy is the highest for Hg ($0.91$~eV/\AA) and the lowest for Ti ($0.21$~eV/\AA); among zz/sta, the interface energy is the highest for Cs ($0.93$~eV/\AA) and the lowest again for Ti ($0.24$~eV/\AA); among ac/str, the interface energy is the highest for Hg and Ag ($0.82$~eV/\AA) and the lowest for Mn ($0.28$~eV/\AA); and among ac/sta, the interface energy is the highest for W ($0.92$~eV/\AA) and the lowest for Ti and Cr ($0.35$~eV/\AA). [Three interfaces, Pb(ac/sta), Tl(ac/sta), and Sr(ac/str) couldn't be fully relaxed.]
Among graphene(zz) edges, the interface energies are nearly always smaller for metallenes with straight edges. Only for Li has zz/sta slightly smaller energy than zz/str (Fig.~S6). A similar trend also holds for graphene(ac) edges, with only a few exceptions. 

For all metals, two-thirds of graphene(zz) interfaces are more stable than graphene(ac) interfaces, and $85$~\%\ of zz/str interfaces are more stable than ac/str interfaces. The better stability of graphene(zz) interfaces stems from both the better average (geometrical) matching with metallene edges, and from the chemically reactive dangling bonds at the zigzag edge \cite{Pekka_2008}. These trends imply that, qualitatively, the energetically most stable interfaces have \emph{geometrically smooth profiles}. Consequently, in what follows, we restrict ourselves to relatively smooth zz/str and zz/sta interfaces, which generally offer lower interface energies (Fig.~S7) and better structural stabilites.

\subsection{Stability}
\begin{figure}[t]
    \centering
    \includegraphics[width=\columnwidth]{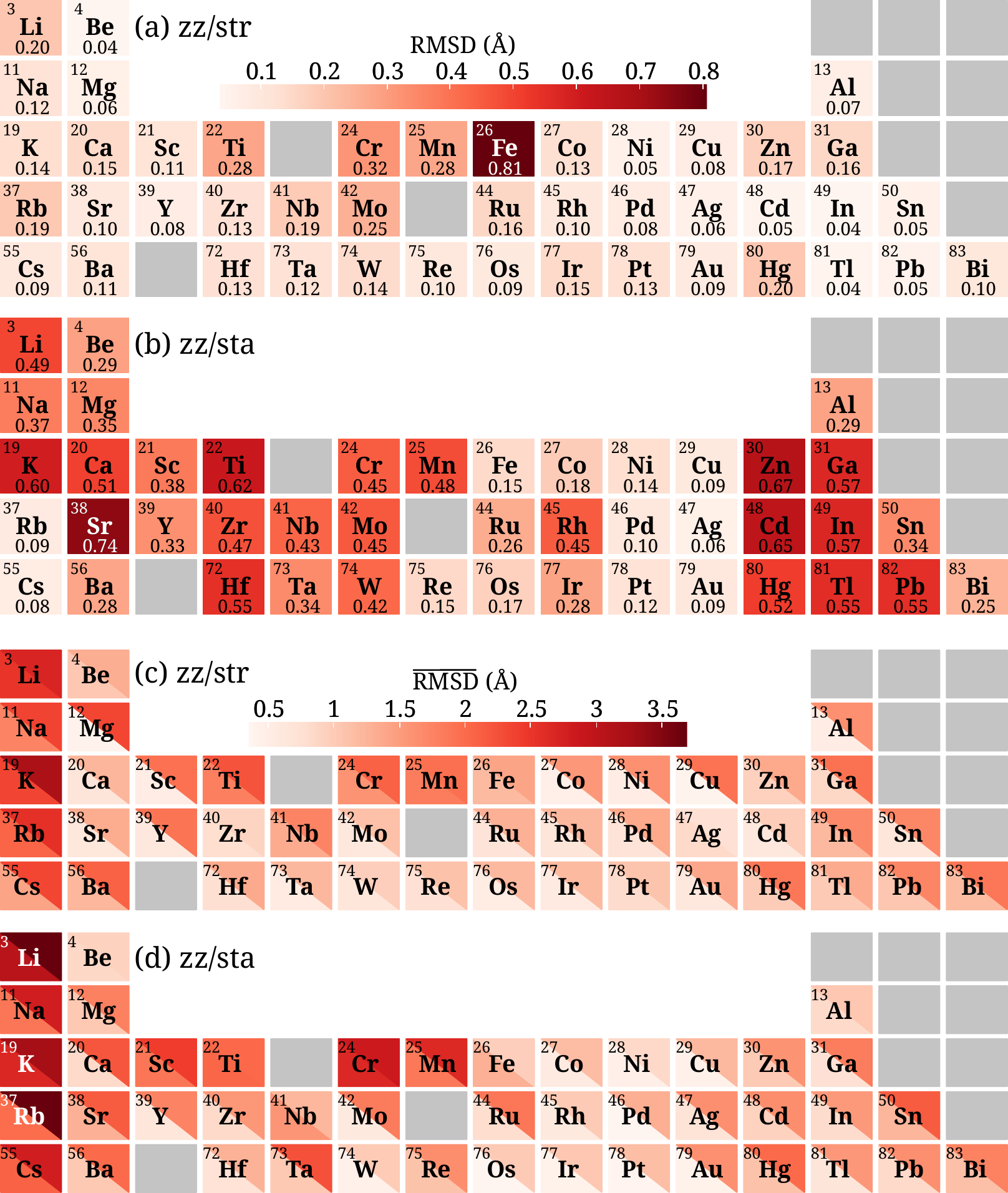}
    \caption{Structural stabilities of graphene(zz)/metallene interfaces. Heat maps of RMSD of (a) zz/str and (b) zz/sta interfaces, comparing initial and relaxed geometries with $\approx 0$\% mismatch.
    Heat maps of average root-mean-square deviation ($\overline{RMSD}$) from MD simulation for (c) zz/str and (d) zz/sta interfaces. In each element box, the top triangle corresponds to $\approx 5$\%\ and the bottom triangle to $\approx 0$\%\ mismatch. For more details, see Fig.~S11.}
    \label{fig:RMSD}
\end{figure}

To complement the energy analysis, we investigate the interface stabilities by reconstruction analysis and MD simulations. During optimization, some of the metallene edges (mostly staggered) got reconstructed or bent compared to the initial structure. We calculated the root-mean-square deviation (RMSD) for each optimization trajectory to quantify these reconstructions. We defined it as
\begin{equation}
    \text{RMSD} = \sqrt{\frac{1}{N}\sum_{i=1}^{N}\delta^2_i},
    \label{eq:rmsd}
\end{equation}
where $\delta_i$ is the displacement of atom $i$ during relaxation and $N$ is the total number of atoms. The zz/str interfaces were optimized very close to their ideal (guessed) shape (Fig.~\ref{fig:RMSD}a). Among them, Be, In, and Tl reconstructed the least, and Fe the most, due to an overall bending of the ribbon (Fig.~S8a). Among zz/sta interfaces, Ag and Cs showed the smallest reconstruction (Fig.~\ref{fig:RMSD}b). In contrast, zz/sta Sr interface got significantly reconstructed: the strong dangling bonds at the zz edge of graphene ripped atoms from the metallene to make metallic chain-decorated graphene edges (Fig.~S8b). The interfaces with staggered edges were reconstructed for RMSD mostly above $0.15$~\AA, straight edges for much less. On average, the reconstructions in graphene(ac) interfaces were $\approx 70$~\%\ larger than in graphene(zz) interfaces (Fig.~S9).

Equipped with this understanding of interface reconstructions, we also tested interface stabilities at room temperature. We conducted ML-based MD simulations for each interface, both at minimum ($\approx 0$\%) and at maximum ($\approx 5$\%) lattice mismatch. The interface stabilities were quantified by averaging the RMSD from Eq.~(\ref{eq:rmsd}) over all time steps (with respect to the relaxed structure),
\begin{equation}
    \overline{\text{RMSD}} = \frac{1}{T}\sum_{t=1}^{T}\text{RMSD}(t),
\end{equation}
where $T$ is the total number of time steps. Thus, $\overline{\text{RMSD}}$ characterizes the dynamical stability of the interface, with a large value indicating greater potential for structural instability. The resulting $\overline{\text{RMSD}}$s follow roughly the trends in reconstruction-RMSDs (Figs.~\ref{fig:RMSD}c and d), but with values some four times larger at room temperature. A larger lattice mismatch causes larger reconstruction and decreases dynamic stability. The large reconstructions at zz/sta interfaces often make the entire interface unstable: many zz/sta interfaces become unstable already after a few picoseconds. These observations for interfaces contrast with the intrinsic stability trends of gas-phase clusters \cite{Janne2019}.

To further validate the ML model, we benchmarked the ML-based MD trajectory of the zz/sta interface of Au against the corresponding DFT-based MD trajectory. For the $1$-ps MD run, the ML and DFT energies agreed to within MAE of $0.88$~eV (Fig.~S10). The atomic forces at each step agreed well both in magnitudes and directions, with the largest average force difference being only $0.8$~meV/\AA\ (less than our force criterion for optimization). The radial distribution functions for Au-Au and Cu-Cu were practically identical. The differences in ML and DFT average bond lengths for each MD step were tiny: MAE was only $0.002$~\AA\ for C-C and $0.009$~\AA\ for Au-C bonds. These small differences indicate that ML is sufficiently accurate and valid also for MD simulations.

To verify that the RMSD definition appropriately captures model dynamics, we examined five representative interfaces at 300 K (Fig.~S12). The analysis indicates that all atoms contribute to maintaining interface stability, such that the calculated RMSD represents the overall structural disorder. Limiting the calculation to atoms near the interface would artificially increase the apparent stability.

\subsection{Microscopic properties}

Finally, we have a detailed look at the interfaces and investigate their charge transfer, atomic defects, and tensile strengths. For this, we focused on selected interfaces in Table~\ref{table:inf}, which cover different parts of the periodic table and are valid candidates for further investigations based on stability analysis. Only Mg(zz/sta) and Ga(zz/sta) raise some concerns, as discussed earlier. In this section, we used DFT, since ML cannot provide information about charge transfer. Although ML can provide tensile strength, we used DFT to ensure consistency within this section.

\subsubsection{Charge transfer}
\begin{figure}[t]
    \centering
    \includegraphics[width=\linewidth]{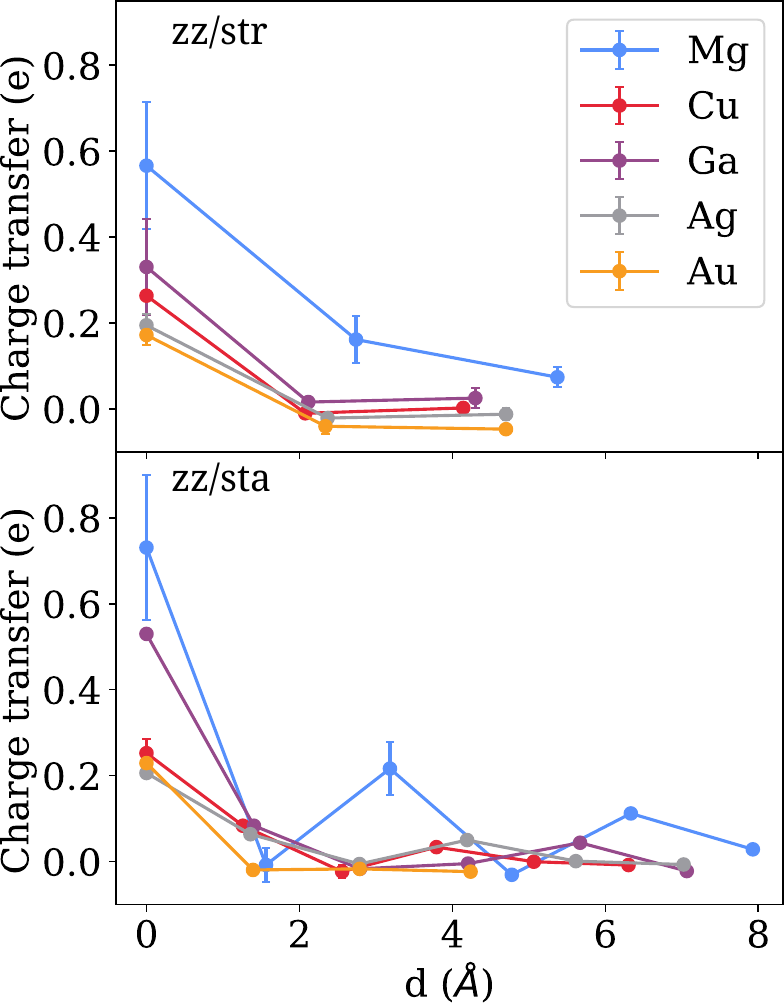}
    \caption{Charge transfer at the interface. The electronic charge transfer per metal atom as a function of distance ($d$) from the interface. (Positive meaning metal atoms lose electrons.) The values are averaged over atoms at roughly the same distance along the interface; the error bar shows the standard deviation of this variation.}
    \label{fig:charge}
\end{figure}

To better understand the electronic structure of interfaces, we performed charge transfer analysis. As expected, the maximum charge transfer occurs right at the interfaces and involves metal atoms that bind directly to carbon (Fig.~\ref{fig:charge}). Away from the interfaces, the charge transfer decreases for all metallenes and goes nearly to zero for metals like Cu and Ag. Charge transfer also varies along the interface, due to the varying bond lengths arising from the incommensurability and lattice mismatch. Significant charge transfer can be observed for bent interfaces, either as the result or the cause of the bending. The coinage metals Cu and Au are illustrative examples of charge transfer that is very even along the interface. Therefore, we can conclude that geometrically smooth interface profiles come with homogeneous charge transfer that contributes positively to stability.

\subsubsection{Defect formation energies}

\begin{figure}
    \centering
    \includegraphics[width=\linewidth]{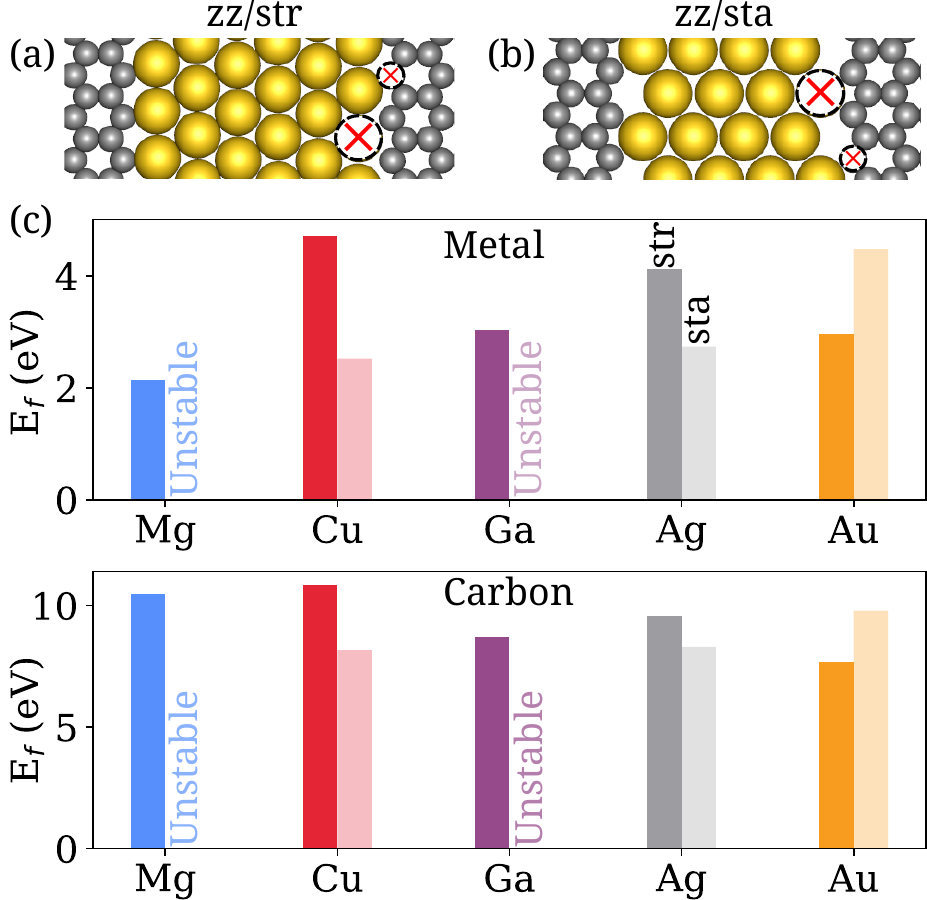}
    \caption{Atom vacancy models for the interfaces. Vacancies (a) at zz/str and (b) zz/sta interfaces. The vacancy candidates were selected based on our charge transfer analysis and marked with red crosses.
    (c) The formation energies of a metal/carbon defects at the interface. Staggered edges are indicated by a brighter color
and straight edges by a darker one.
    }
    \label{fig:defect}
\end{figure}

Apart from varying lattice mismatch, understanding realistic micro-structural features of graphene/metallene interfaces requires modeling defects, such as vacancies. Defects are important for stability and lateral expansion \cite{koskinen_2021}. Here, we investigated single metal and carbon defects at the interfaces (Figs.~\ref{fig:defect}a and b). We removed atoms with the largest charge transfer and calculated the formation energy $E_\text{f}$ for a single  defect as
\begin{equation}
    E_\text{f} = E_\text{def} - E_\text{pris} + \mu_\text{free},
\end{equation}
where $E_\text{def}$ is energy of the defected system, $E_\text{pris}$ is energy of the pristine system, and $\mu_\text{free}$ is chemical potential of the free atom (Fig.~\ref{fig:defect}c).

The resulting metal vacancy formation energies range between $2.1$--$4.7$ eV. Carbon defect formation energies are much larger and range between $7.7$--$10.9$ eV.
As concluded from stability analysis, Ga(zz/sta) and Mg(zz/sta) are omitted. They got reconstructed or deformed too much in the pristine structure, so that introducing defects made them unstable to an extent that rendered the calculation of defect formation energies meaningless.

In alignment with energy and stability analysis, the results show that, compared to zz/sta, zz/str interfaces with smoother profiles are more stable against metal defect formation. The formation energies of carbon defects have much less variation. The small variation is understandable because carbons bind the strongest to their neighboring carbon environment, which remains fairly similar irrespective of the metallene.

\subsubsection{Tensile strength}

We examined the mechanical stability of interfaces by calculating their tensile strengths. They were calculated as the maximum mechanical tensile stress along the $x$-direction (perpendicular to the interface) that the structure could withstand before breaking bonds. The interface stress (range 1-8\%) was applied by straining the entire simulation cell in the $x$-direction (expanding cell volume). The stress tensor was calculated in Voigt order, and the $xx$ component was extracted.

\begin{figure}[h]
    \centering
    \includegraphics[width=\linewidth]{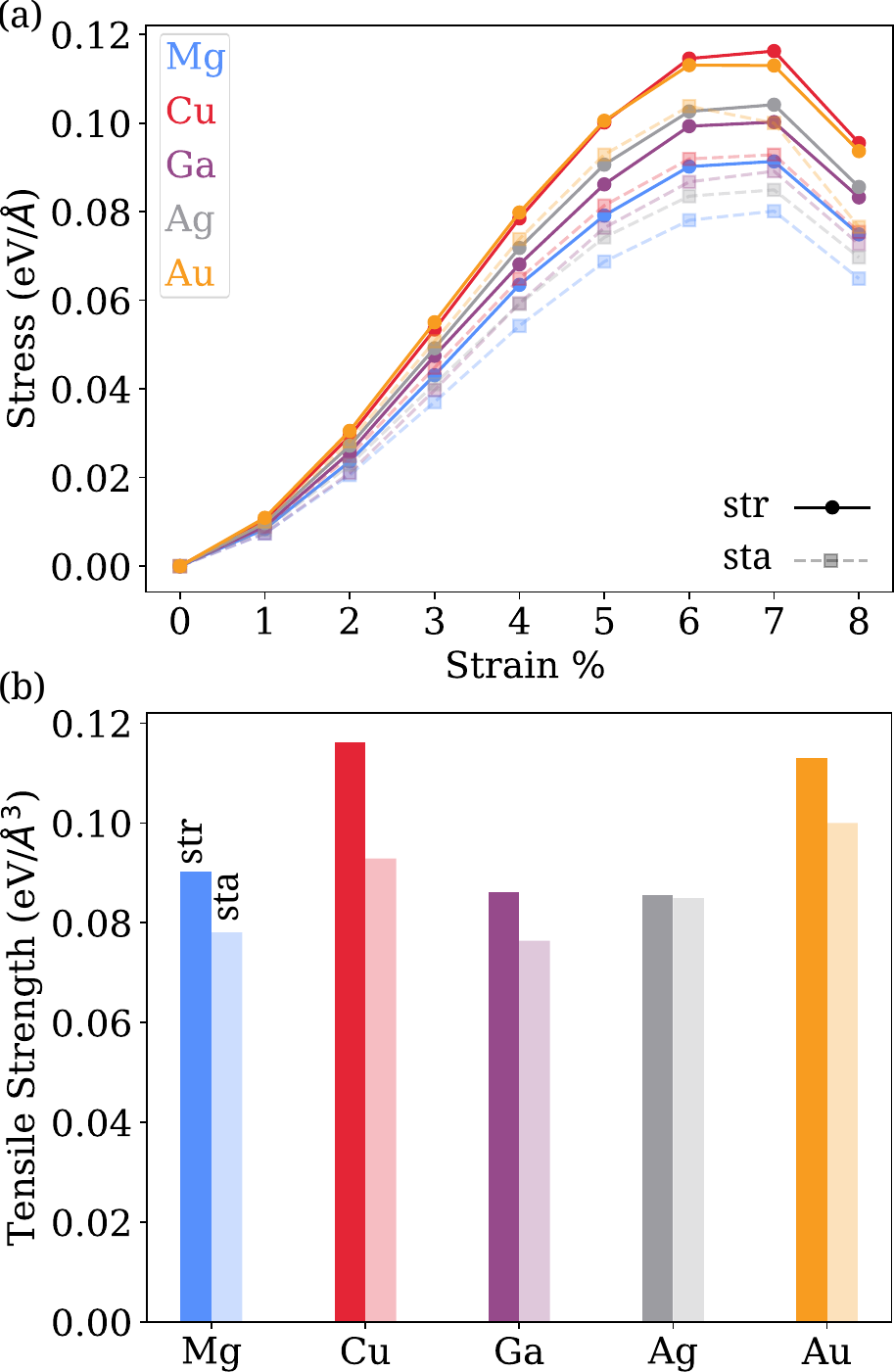}
    \caption{Mechanical stability of interfaces. (a) Stress vs. strain in selected zz/metallene interfaces. Metals are denoted by colors, zz/str by circles and solid lines, and zz/sta by squares and dashed lines. (b) Tensile strengths for the selected zz/str and zz/sta interfaces. Staggered edges are indicated by a brighter color
and straight edges by a darker one. }
    \label{fig:tensile}
\end{figure}

Overall, the stress curves are consistent across all metallenes (Fig.~\ref{fig:tensile}a). All interfaces are stable up to $7$~\%\ strain and have tensile strengths larger than $0.07$ eV/\AA\ (Fig.~\ref{fig:tensile}b). Still, stresses at zz/sta interfaces increase more slowly upon increasing strain than at zz/str interfaces. In particular, the zz/str interfaces always have larger tensile strengths and greater stabilities than zz/sta interfaces. This trend again agrees with stability analysis (Sec. 3.2). Among transition metals, Cu and Au have similar tensile strengths, reflecting a slightly better stability than other interfaces.

\section*{Conclusions}
In this article, we studied the structural, energetic, dynamic, and electronic properties of lateral graphene-metallene interfaces. Our main computational tool was DFT, but we also demonstrated that the off-the-shelf ML model MatterSim could reliably predict the energetic properties of graphene–metallene interfaces without any additional fine-tuning. We validated the ML model against DFT calculations for five representative metals (Mg, Cu, Au, Ag, and Ga) using a few selected interface configurations. A fast and reliable ML model enabled a systematic investigation of interface geometries and energies for $45$ metals and four different interface profiles involving zigzag and armchair edges of graphene and straight and staggered edges of metallene. Although the MatterSim exhibits strong agreement with DFT across a wide range of structural and energetic properties, it is well recognized that transferable ML models may face challenges in accurately capturing charge transfer and defect energetics. Consequently, despite the overall success of the ML model, we naturally had to resort to DFT for microscopic interface properties where first-principles methods remain indispensable.

We found that the most stable interfaces---both energetically and mechanically, both in pristine and in defective states---are zz/str interfaces. Systematic analysis indicated an even more general correlation: interfaces with geometrically smooth profiles are significantly more stable both energetically and dynamically. We focused on zigzag edges of graphene, but ML results also support this conclusion for armchair edges. Regarding electronic structure, the charge transfer was the largest near the interface and reduced greatly as the distance to the interface increased. Lattice mismatch and concomitant incommensurability also created inhomogeneities in the charge transfer. These variations imply that the chemical behavior of metal atoms near interfaces can depend dramatically on position.

We conclude that the most stable interfaces are found for transition metals with smooth profiles. These findings are aligned with the experimental fabrication methods for synthesizing metallenes \cite{Ta2021, Fengzhu2025, Binqi2025, TANG20191454} such as Cr \cite{Cr2020}, Fe \cite{Fe2014}, Zr \cite{Zr2025}, Mo \cite{Mo2018}, and Sn \cite{Sn2020} using atomically focused e-beam sculpting inside graphene or other 2D template pores. Although we have predicted which graphene-metallene interfaces will be more stable, addressing the actual, experimental route to reach them ultimately is beyond our scope.

We hope that both our specific results and especially the rule-of-thumb trends across different interfaces will guide both computational and experimental research of metallenes involving interfaces. It is plausible that similar trends could be extracted for other two-dimensional stabilizing templates, such as hexagonal boron nitride (BN) and transition metal dichalcogenides (\ce{MoS2}, \ce{WS2}, etc.). Particularly when their edge profile approximates the zigzag configuration of graphene, thereby satisfying geometric smoothness conditions that favor stability.
However, our results demonstrate that geometrical details are essential; the trends must be investigated for each template material separately.

\section*{Author contributions}
M. Bagheri: calculations, investigation, validation, formal analysis, visualization, methodology, writing – original draft; P. Koskinen: conceptualization, resources, methodology, supervision, funding acquisition, project administration, writing – review \& editing.

\section*{Conflicts of interest}
There are no conflicts to declare.

\section*{Data availability}

The optimized interfaces and their energies from calculations are saved in a database available at \url{https://doi.org/10.5281/zenodo.15720866}.
Other data and scripts used in this study are available from the first author upon reasonable request.

\section*{Acknowledgements}

We acknowledge the Jane and Aatos Erkko Foundation for funding (project EcoMet) and the Finnish Grid and Cloud Infrastructure (FGCI) and CSC—IT Center for Science for computational resources. 



\balance


\bibliography{refs} 
\bibliographystyle{rsc} 
\end{document}